\documentclass[aps,prx,twosuperscriptaddress,floatfix]{revtex4}

\usepackage{graphicx,color}
\usepackage{dcolumn}
\usepackage{bm}
\usepackage{stmaryrd}
\usepackage{latexsym}
\usepackage{amssymb}
\usepackage{amsfonts}
\usepackage{amsmath}

\begin{document}

\title{Exotic electronic states in the world of flat bands: from theory to material }

\author{Zheng Liu}
\affiliation{Department of Materials Science and Engineering, University of Utah, Salt Lake City, UT 84112, USA}

\author{Feng Liu}
\affiliation{Department of Materials Science and Engineering, University of Utah, Salt Lake City, UT 84112, USA}
\affiliation{Collaborative Innovation Center of Quantum Matter, Beijing 100084, China}

\author{Yong-Shi Wu}
\affiliation{State Key Laboratory of Surface Physics and Department of Physics, Fudan University, Shanghai 200433, China}
\affiliation{Department of Physics and Astronomy, University of Utah, Salt Lake City, UT 84112, USA}

\date{\today}


\begin{abstract}
It has long been noticed that special lattices contain single-electron flat bands (FB) without any dispersion. Since the kinetic energy of electrons is quenched in the FB, this highly degenerate energy level becomes an ideal platform to achieve strongly correlated electronic states, such as magnetism, superconductivity and Wigner crystal. Recently, the FB has attracted increasing interests, because of the possibility to go beyond the conventional symmetry-breaking phases, towards topologically ordered phases, such as lattice versions of fractional quantum Hall states. This article reviews different aspects of FBs in a nutshell. Starting from the standard band theory, we aim to bridge the frontier of FBs with the textbook solid-state physics. Then, based on concrete examples, we show the common origin of FBs in terms of destructive interference, and discuss various many-body phases associated with such a singular band structure. In the end, we demonstrate real FBs in quantum frustrated materials and organometallic frameworks.
\end{abstract}

\maketitle

\section{Introduction and scope}

A central component to understand the motion of electrons in a crystalline solid is the electronic band structure, which defines the relationship between an electron's energy and its momentum. The electronic band theory has laid down the theoretical foundation for electronic devices, leading to the birth of modern information technology.

Normally, electrons in a crystal (the so-called Bloch electrons) move just like free electrons, except for a different effective mass as defined by the band dispersion. In certain lattices, the band dispersion can drastically differ from that of a free electron. One extreme instance is the linearly dispersive bands in a 2D hexagonal lattice, which makes the Bloch electrons massless \cite{RMP09GeimGraphene,RMP11Geim}. There can also be lattices lifting the electrons to the other end of the mass spectrum, i.e. arbitrarily large effective mass. In 1991, Mielke \cite{JPA91Mielke,JPA91Mielke2,JPA92Mielke} found that a special class of lattices, mathematically known as the line graphs, contain electronic bands which are completely flat. The existence of flat bands (FB) was later extended to other families of conceptual lattices \cite{PRL92Tasaki}. Since the kinetic energy of electron is quenched in the FB, the Coulomb interaction becomes critical, giving rise to various exotic many-body states, such as ferromagnetism \cite{JPA91Mielke,JPA91Mielke2,JPA92Mielke,PRL92Tasaki,zhang2010proposed}, superconductivity \cite{PhysC07BCSFB} and Wigner crystal \cite{PRL07Wu,PRB08Wu}.

Bloch electrons may additionally have an anomalous velocity transverse to their normal motion as observed by Hall more than a century ago, known as the (anomalous) Hall effect. Such an anomalous motion has subtle connections with the Berry phase of Bloch wavefunction in the momentum space \cite{RMP10XiaoDi}. After realizing that the global configuration of Bloch wavefunction can be viewed as a topological structure, the concept of topology is introduced to classify bands using the topological numbers, e.g. Chern number \cite{PRL82TKNN,PRB88HaldaneModel,PRB06QiWuZhang} and $Z_2$ number \cite{PRL05KaneMele,RMP10Hasan,RMP11Qi}, which expands the band theory to new territories.

In 2011, three research groups independently proposed the idea of introducing nontrivial topology to the FB to achieve strongly-correlated topological states \cite{PRL11Tang,PRL11Sun,PRL11Neupert}. This proposal was motivated by an important question frequently asked by condensed matter physicists: Can the quantum Hall (QH) effects on Landau levels (LL) be generalized to Bloch bands in the absence of an external magnetic field. The proposed topological FB is just like a counterpart of the LL, characterized by a Chern number equal to 1 \cite{PRL82TKNN}. It has been further shown by numerical simulations that such FBs support QH-like states, exhibiting not only the integer QH effects, but also the fractional QH (FQH) effects \cite{PRL11Neupert,NatCom11Sheng,PRX11FCI}. More importantly, considering the energy scales in lattices, the FQH effects -- such as fractionalization and entanglement -- can be realized in a much higher temperature in the FB than in the LL \cite{PRL11Tang}, charting a revolutionary route towards quantum computation \cite{RMP08QuantComp}.

A few nice reviews have already been written on different aspects of the FB, such as ferromagnetism \cite{PTP98Tasaki} and FQH effects \cite{CRP13ReviewFB,Liu13ReviewFB}. We intend not to repeat these previous efforts. Instead, we try to introduce this field in a nutshell from a different angle - starting from the standard band theory, and then expanding discussions based on concrete examples, including discussions on material realizations of FBs. We target at those readers coming from different disciplines, e.g. semiconductor physics, materials science or even chemistry, rather than being experts on many-body physics or FQH physics. We expect that this review article will provide an easy-to-understand flavor for the basic ideas of FBs, and possibly motivates wider interests in this topic. The article is organized as follows. Sec. \ref{sec:FB} contains a survey of FB lattices. Sec. \ref{sec:wannier} explains the common origin of FBs in terms of the localized eigenstates. Sec. \ref{sec:compare} compares the FB with the LL. Sec. \ref{sec:Manybody} describes various many-body phases that may emerge in the FB systems (with interactions among electrons taken into account). Sec. \ref{sec:phasediagram} constructs the phase diagram of a practical FB system. Sec. \ref{sec:material} discusses possible material realizations of FBs. Sec. \ref{sec:end} concludes this article. We limit our discussions to the most well-studied 2D FBs, and purposely avoid invoking abstract mathematical derivations in this article. With the intuitive pictures provided here, we encourage the readers to work on more specialized literature for the rigorous proof \cite{PTP98Tasaki,CRP13ReviewFB,Liu13ReviewFB}. Also, we primarily focus on realizing electronic FBs in solid-state materials using the charge degree of freedom. The readers interested in bosonic FBs in ultracold atom systems, or magnon FBs in frustrated antiferromagnets are referred to \cite{PRL13FBOpticalFlux,PRL12Norman,PRL13Norman,derzhko2007universal}.

\section{How to make a band flat?} \label{sec:FB}

We start from electron hopping on a general lattice:

\begin{eqnarray} \label{eq:hopping}
H_{hop}&=&\sum_{i,j,a,b} t_{ia,jb} c_{i,a}^\dag c_{j,b},
\end{eqnarray}
where $i$,$j$ index unit cells, $a$,$b$ label different orbitals within one unit cell, and $\{t_{ia,jb}\}$ are the hopping matrix elements.

With the aid of Bloch theorem, $H_{hop}$ can be transformed to the momentum space as:

\begin{eqnarray}
h_\textbf{k}=\sum_{\textbf{k},a,b} c_{\textbf{k},a}^\dag h_{ab} (k) c_{\textbf{k},b}
\end{eqnarray}
Solving the eigenvalue problem of $h(\textbf{k})$ gives the Bloch bands $\{\varepsilon_{n\textbf{k}}\}$ and Bloch states $\{\psi_{n\textbf{k}}\}$.

As expressed by the name ``flat band'', we aim at finding a very singular band structure, in which one or more Bloch bands are completely dispersionless, i.e. $\varepsilon_{n\textbf{k}}=const$. A trivial solution is to set $t_{ia,jb}=0$, which corresponds to the isolated atomic limit. Those narrow bands commonly existing in heavy fermion compounds arise from this picture, which is not the focus of the present article. Besides this trivial solution, nontrivial solutions do exist, which retain the hopping dynamics throughout the lattice, whereas making the net hopping vanish. This condition is nothing but destructive interference. Theoretically, one can simply obtain one nontrivial solution by performing an adiabatic transformation: $\tilde{h}(\textbf{k})= h(\textbf{k})/ \varepsilon_{n\textbf{k}}$ \cite{CRP13ReviewFB,Liu13ReviewFB}. Fourier transforming $\tilde{h}(\textbf{k})$ back to the real space then defines the new hopping matrix elements, which creates complete destructive interference and makes the $n$th band dispersionless. This method in principle flattens any individual Bloch band, but usually results in long-range hopping elements, which are unrealistic in real materials.

Is it possible to create the destructive interference more naturally? We review a list of approaches below.

\subsection{Line graph} \label{sec:line}

\begin{figure}[tbp]
\begin{center}
\tabcolsep=-0.5cm
\begin{tabular}{c}
\includegraphics[width=8.5cm]{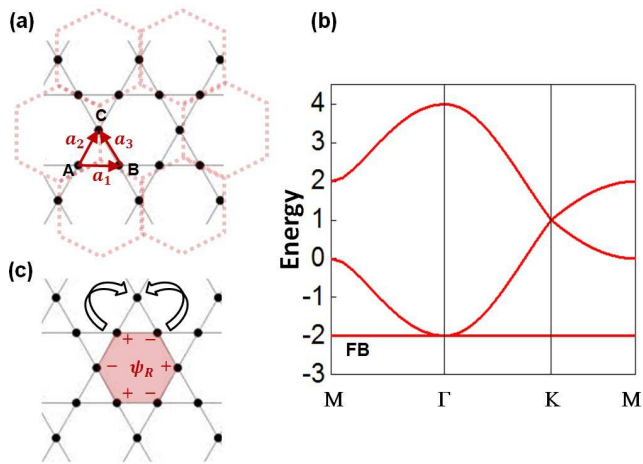}
\end{tabular}
\end{center}
\caption{(a) The kagaome lattice (black solid lines) as the line graph of a honeycomb  lattice (red dashed lines). (b) Band structure from Eq. (\ref{eq:Hkagome}) with $t=1$ (c) A localized eigenstate of the FB and the destructive interference. }\label{fig:kagome}
\end{figure}

As pointed out by Mielke \cite{JPA91Mielke,JPA91Mielke2,JPA92Mielke} , the destructive interference is a common feature of a special class of lattices, mathematically known as the line graphs (We will see the interference pattern explicitly in Sec.\ref{sec:wannier}). Line graph is a specific geometry transformation of an original lattice, which can be roughly viewed as a bond-site exchange. Rigorously, given the original lattice, the line graph is obtained by drawing a site for each of the bonds in the original lattice, and then if two bonds in the original lattice share a common site, these two corresponding sites in the line graph are connected with a bond.

As a typical example, consider a 2D honeycomb lattice. For this case, the construction of line graph as described above results in the geometry of corner-sharing triangles [Fig. \ref{fig:kagome}(a)]. This lattice is commonly called the ¡°kagome¡± lattice, getting the name from Japanese for a weave pattern of basket. The Bravais lattice is hexagonal. Within one unit cell, there are three inequivalent sites $A,B,C$ [Fig. \ref{fig:kagome}(a)]. We consider a single orbital on each site, i.e. $a,b=A,B,C$ with respect to Eq. (\ref{eq:hopping}),  and set a common hopping amplitude $t$ for all the nearest neighbor (NN) bonds.  The Bloch Hamiltonian is then:

\begin{eqnarray} \label{eq:Hkagome}
  h(\textbf{k})=
    \left(
     \begin{array}{ccc}
       0 & 2t \cos k_1 & 2t \cos k_2 \\
       2t \cos k_1 & 0 & 2t \cos k_3 \\
       2t \cos k_2 & 2t \cos k_3 & 0 \\
     \end{array}
  \right),
\end{eqnarray}
where $k_n=\textbf{k} \cdot \textbf{a}_n$ with the definition of $\textbf{a}_n$ given in Fig. \ref{fig:kagome}(a). The eigenvalues of this $3 \times 3$ matrix form one flat band $\varepsilon_1 = -2t$, and two dispersive bands $\varepsilon_{2,3} = t [1\pm \sqrt{(4(\cos^2 k_1+\cos^2  k_2 +\cos^2 k_3)-3}]$ [see Fig. \ref{fig:kagome}(b)]. The Bloch state of the flat band is:

\begin{eqnarray}\label{eq:KFB}
\psi_{FB,\textbf{k}}^\dag=\sin k_3 c_{\textbf{k},A}^\dag-\sin k_2 c_{\textbf{k},B}^\dag + \sin k_1 c_{\textbf{k},C}^\dag
\end{eqnarray}

Other important examples of line graphs include the checkerboard lattice \cite{PRL11Sun}, which is transformed from the simple square lattice. Its equivalent representation in 3D is the pyrochlore lattice, consisting of corner-sharing tetrahedral. Many decorated lattices can be viewed as partial line graphs, by applying the transformation to one set of the sublattices \cite{JPSP05partialLG}.

\subsection{Cell construction} \label{sec:cell}

\begin{figure}[tbp]
\begin{center}
\tabcolsep=-0.5cm
\begin{tabular}{c}
\includegraphics[width=8.5cm]{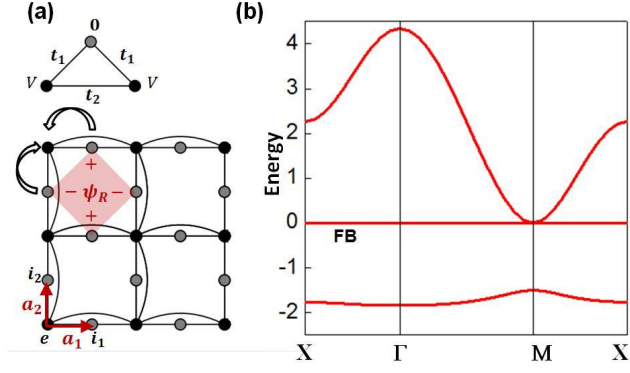}
\end{tabular}
\end{center}
\caption{(a) The side-centered square lattice (bottom) constructed by assembling triangular cells (top). The shaded region in the square lattice is a localized eigenstate of the FB. The arrows indicate the destructive interference. (b) Band structure from Eq. (\ref{eq:Hsquare}) with $t_1=1$, $t_2=0.5$ and $V=0.5$ }\label{fig:square}
\end{figure}

Another systematic way to construct FBs was proposed by Tasaki called ``cell construction'' \cite{PTP98Tasaki}. The construction process starts from an elemental cell consisting of a single internal site and two or more external sites. These cells are then assembled to form a lattice by sharing the external sites. For example, let us consider the smallest cell, a triangle. A typical 2D assembling is shown in Fig. \ref{fig:square}a. It forms a side-centered square (SCS) lattice just like the $CuO_2$ plane in cuprate superconductors. Within one unit cell, there are two inequivalent internal sites and one external site. We consider a single orbital on each site, i.e. $a,b=i_1,i_2,e$ with respect to Eq. (\ref{eq:hopping}), and assign two different hopping amplitudes $t_1$, $t_2$ for the hopping processes as shown in Fig. \ref{fig:square}(a). The on-site energy between the external and internal sites can have a difference denoted by $V$. The corresponding Bloch Hamiltonian is then :

\begin{eqnarray} \label{eq:Hsquare}
  h(\textbf{k})=
    \left(
     \begin{array}{ccc}
       0 & 0 & 2t_1 \cos k_1 \\
       0 & 0 & 2t_1 \cos k_2 \\
       2t_1 \cos k_1 & 2t_1 \cos k_2 & 2\Lambda_\textbf{k} \\
     \end{array}
  \right)
\end{eqnarray}
where $k_n=\textbf{k} \cdot a_n$ with the definition of $\textbf{a}_n$ given in Fig. \ref{fig:square}(a), and  $\Lambda_\textbf{k}=t_2 (\cos 2k_1+\cos 2k_2)+\frac{V}{2}$. The eigenvalues of this $3 \times 3$ matrix consist of one FB: $\varepsilon_1=0$ and two dispersive bands: $\varepsilon_{2,3}=\Lambda_\textbf{k}\pm\sqrt{\Lambda_\textbf{k}^2+2t_1^2 (2+\cos 2k_1+\cos 2k_2)}$ [See Fig. \ref{fig:square}(b)]. The Bloch state of the FB takes the form:

\begin{eqnarray}\label{eq:SFB}
\psi_{FB,\textbf{k}}^\dag=\cos k_2 c_{\textbf{k},i_1}^\dag-\cos k_1 c_{\textbf{k},i_2}^\dag
\end{eqnarray}

\subsection{Orbital selection}

\begin{figure}[tbp]
\begin{center}
\tabcolsep=-0.5cm
\begin{tabular}{c}
\includegraphics[width=8.5cm]{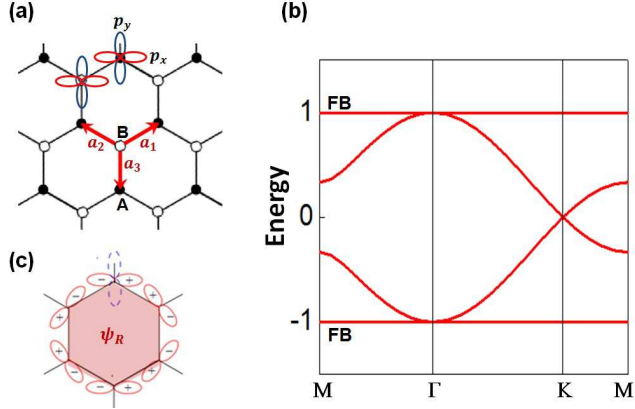}
\end{tabular}
\end{center}
\caption{(a) The $p_{x,y}$-orbital honeycomb lattice. (b)  The band structure from Eq. (\ref{eq:Hpxy}) with $t_\sigma=1$ (c) The localized eigenstate of the FB. The electron cannot leak outside if $\pi$-type hopping is set to zero.}\label{fig:honeycomb}
\end{figure}

 In addition to designing the lattice geometry, FBs can also obtained by exploiting the orbital freedom.  Wu et al. \cite{PRL07Wu,PRB08Wu} proposed a $p_{x,y}$-orbital counterpart of graphene (Fig. \ref{fig:honeycomb}a), by considering that each of the two sites in a honeycomb lattice contains two orbitals $p_x$ and $p_y$, i.e.  $a,b=p_{Ax}, p_{Ay}, p_{Bx}, p_{By}$ with respect to Eq. (\ref{eq:hopping}). We will denote this lattice by XYH in the remainder of this article.  By convention, the hopping processes between the $p_{x,y}$-orbitals can be classified into the $\sigma$-type (head to tail) and the $\pi$-type (shoulder by shoulder). We assume that the $\pi$-type hopping is negligible. The Bloch Hamiltonian is written as:

\begin{eqnarray} \label{eq:Hpxy}
&h&(\textbf{k})= \\ \nonumber
&t_\sigma& \left(
  \begin{array}{cccc}
    0 & 0 & \frac{3}{4}(e^{ik_1}+e^{ik_2}) & \frac{\sqrt3}{4}(e^{ik_1}-e^{ik_2}) \\
    0 & 0 & \frac{\sqrt3}{4}(e^{ik_1}-e^{ik_2}) & \frac{1}{4}(e^{ik_1}+e^{ik_2})+e^{ik_3}) \\
    h.c. &  & 0 & 0 \\
     &  & 0 & 0 \\
  \end{array}
\right),
\end{eqnarray}
where $k_n=\textbf{k} \cdot \textbf{a}_n$ and the definition of $\textbf{a}_n$ is given in Fig. \ref{fig:honeycomb}c . The eigenvalues of this $4 \times 4$ matrix consist of two FBs $\varepsilon_{1,4}=\pm \frac{3 t_\sigma}{2}$ and two dispersive bands: $\varepsilon_{2,3}=\pm \frac{t_\sigma}{2}\sqrt{3+2(\cos k_1 + \cos k_2 + \cos k_3)}$ [see Fig. \ref{fig:honeycomb}(b)]. The Bloch states of the two FBs take forms:

\begin{eqnarray}\label{eq:HFB}
\psi_{FB,\textbf{k}}^\dag=\frac{1}{\sqrt3}[f^*_{23}(\textbf{k})-f^*_{31}(\textbf{k})] c^\dag_{\textbf{k},p_{Ax}}-f^*_{12}(\textbf{k})c^\dag_{\textbf{k},p_{Ay}}  \\ \nonumber \pm \frac{1}{\sqrt3}[f_{23}(\textbf{k})-f_{31}(\textbf{k})]c^\dag_{\textbf{k},p_{Bx}} \mp f_{12}(\textbf{k})c^\dag_{\textbf{k},p_{By}}
\end{eqnarray}
where $f_{nm}=e^{ik_n}-e^{ik_m}$.

Note that this model should not be confused with the $\sigma$-bands in graphene, where the $s$-orbital together with the $p_{xy}$-orbitals form a six-dimensional basis. Here, the $s$-orbital is not involved. Sun et al. \cite{PRL11Sun} have made a similar proposal based on a $p_{x,y}$+$d_{x^2-y^2}$ square lattice. In general, by enforcing special conditions for the hopping elements, destructive interference can be created. The concern is whether these special conditions can be met in real materials.

\section{Localized eigenstates of flat band} \label{sec:wannier}

We have mentioned that the origin of FBs is destructive interference. We can see this feature more clearly by constructing the localized eigenstates of FBs.

Firstly, note that because of the energy degeneracy, any linear combination of the FB Bloch states is still an eigenstate. Particularly, the Fourier transformation of the FB Bloch states should be eigenstates. For example, let us calculate the Fourier transformation of Eq. (\ref{eq:KFB}):

\begin{eqnarray}\label{eq:wannierkagome}
\psi_{FB,\textbf{R}}^\dag=\emph{N} \int_{BZ} d\textbf{k} e^{-i\textbf{k}\cdot\textbf{R}} \psi_{FB,\textbf{k}}^\dag=\frac{1}{\sqrt6}\sum_{a=1}^6 (-1)^a c_a^\dag
\end{eqnarray}
where $\emph{N}$ is a normalization constant and $a$ runs over the six vertices of a hexagon centered at the chosen \textbf{R}. As shown in Fig. \ref{fig:kagome}(c), $\psi_{FB,\textbf{R}}$ takes the form of a localized hexagonal plaquette. The wavefunction amplitude alternates its sign around the six vertices. Consequently, the net hopping out of the plaquette vanishes, because the hopping originating from the two adjacent vertices cancels each other. This kind of destructive interference effectively traps the electron within the plaquette. The electron thus appears to have a quenched kinetic energy, as dictated by the FB.

One can view $\psi_{FB,\textbf{R}}$ as a Wannier-like function, but be aware that $\psi_{FB,\textbf{k}}$ given by Eq. (\ref{eq:KFB}) is not normalized, so the set of $\{\psi_{FB,\textbf{R}}\}$ is neither orthogonal nor linear-independent. One can immediately verify the following relation:

\begin{eqnarray}\label{eq:sumFB}
\sum_\textbf{R} \psi_{FB,\textbf{R}}=0
\end{eqnarray}
Considering each unit cell contains one hexagonal plaquette, we have only $(N_c-1)$ independent states, where $N_c$ is the number of unit cells. As pointed out by Bergman et al. \cite{PRB08BergmanTouch}, two more extended eigenstates exist for the FB, which in total gives $N_c+1$ eigenstates at the same energy. The extra state manifests in a touching point with the dispersive band.

The ``self-trapped'' localized eigentstates is a common feature of all the FB models. The localized eigenstates of the SCS lattice take the form of squares as shown in Fig. \ref{fig:square}(a). For the XYH lattice, $\{\psi_{FB,\textbf{R}}\}$ form hexagonal plaquettes with the $p$-orbital aligning along the tangential direction [Fig. \ref{fig:honeycomb}(c)]. Since we neglect the $\pi$-type hopping, this configuration prevents the electron from hopping outside.

The destructive interference can be destroyed by additional hopping terms, such as the next-NN hopping for the kagome and SCS lattices, and the $\pi$-type hopping for the XYH lattice, which will then result in dispersion of the FB. These factors should be properly taken into account for material realizatoin.

\section{Comparisons between flat band and Landau level}\label{sec:compare}
 Another type of ``FB'' that has been extensively studied since 1980s, is the LL (See, for example, Sec. 9.6 in \cite{PeterCardona}). The LLs arise from quantized cyclotron motions of free 2D electron gas subjected to a strong perpendicular magnetic field. They form a set of equally spaced spectrum:
\begin{eqnarray}
E_n = (n + \frac{1}{2})\hbar\omega_c , n=0,1,2,...
\end{eqnarray}
where $\omega_c=eB/mc$ is the cyclotron frequency. It is fair to call the LL a ``FB``, because it has a constant energy, independent of momentum. However, the origin of flatness is different. For a lattice FB as discussed in the previous Section, the flatness is achieved by trapping the electrons within localized plaquettes via destructive interference. For a LL, the flatness is achieved by the magnetic field, which ``traps'' the electrons by driving them to do cyclotron motions. Vidal et al. introduced the notation of Aharonov-Bohm cage to describe this effect \cite{vidal1998aharonov}.

There are interesting connections between the localized eigenstates of FBs and the quantized cyclotron orbitals of LLs. By choosing the symmetric gauge, the eigenfunction of the lowest Landau level has an exponentially localized form (See, for example, Ch. 7 in \cite{XiaogangWen}):
\begin{eqnarray}
\psi_{LL}=\exp(-|\textbf{r}|^2/4l_B^2),
\end{eqnarray}
where $l_B=(\hbar/m\omega_c)^{1/2}$ is the magnetic length. When translated to different centers in the plane, it picks up an additional phase from the magnetic vector potential:
\begin{eqnarray}
\psi_{LL,\textbf{R}}=\exp[-|\textbf{r-R}|^2/4l_B^2+i(\hat{z}\cdot\textbf{R}\times\textbf{r})]
\end{eqnarray}
One can view the set of $\{\psi_{LL,\textbf{R}}\}$ as a counterpart of $\{\psi_{FB,\textbf{R}}\}$. A subtle point is that just like Eq.(\ref{eq:sumFB}), $\{\psi_{LL,\textbf{R}}\}$ does not form a complete or linearly independent basis either. One can verify that on the sites of a magnetic lattice with one flux quantum per unit cell, i.e. $\textbf{R}=\sqrt{2\pi}l_B(n_1,n_2)$,  $\{\psi_{LL,\textbf{R}}\}$ obeys the following relation \cite{JPC84ThoulessWannier}:
\begin{eqnarray}
\sum_{\textbf{R}} (-1)^{n_1+n_2+n_1n_2} \psi_{LL,\textbf{R}}=0
\end{eqnarray}

It is known that the obstruction of constructing a complete set of localized functions for the LL has a deep topological origin \cite{JPC84ThoulessWannier}. If a complete set of localized functions could be constructed, the corresponding Bloch functions,
\begin{eqnarray}
\psi_\textbf{k}=\sum_\textbf{R} \psi_\textbf{R} \exp(i\textbf{k}\cdot\textbf{R})
\end{eqnarray}
 would be analytic and single-valued in the whole Brillouin zone (BZ). In opposite, when we find that such a set can not be constructed, it indicates that the Bloch functions have singularities. Usually, such singularity implies nontrivial topology formed by the Bloch functions, such as vortices. Mathematically, the topology of LLs is characterized by a Chern number $c=1$ \cite{PRL82TKNN}. The definition is given by:
\begin{equation}
c=\frac{1}{2\pi i}\int_{BZ} d^2\textbf{k} F_{12}(\textbf{k}),
\end{equation}
where the Berry curvature $F_{12}(\textbf{k})$ is given by:

\begin{eqnarray} \label{eq:berryconn}
F_{12}&=&\partial_1 A_2(\textbf{k})-\partial_2 A_1(\textbf{k})\nonumber\\
A_i&=&\langle \psi_\textbf{k}|\partial _i|\psi_\textbf{k}\rangle,
\end{eqnarray}
where the derivative $\partial_i$ stands for $\partial / \partial_{k_i}$. The Chern number describes the Berry phase accumulation, or winding of Bloch functions in the whole BZ, which physically manifests in the quantized Hall conductance.

With this picture in mind, let us turn back to the FBs to see a deeper connection between the LL and the FB in terms of topology. First note that $\psi_{FB,\textbf{k}}$ indeed contains a singular point. It vanishes at $(0,0)$ for the kagome lattice [Eq.(\ref{eq:KFB})] and the XYH lattice [Eq.(\ref{eq:HFB})], and at $(\frac{\pi}{2}, \frac{\pi}{2})$ for the SCS lattice [Eq.(\ref{eq:SFB})]. The 3-component Bloch functions [Eqs.(\ref{eq:KFB}) and (\ref{eq:SFB})] can be directly visualized as vectors. It is convenient to expand them around the singular point to the first order and then see them winding by $2\pi$ around the singular point. This can be related to a $2\pi$ Berry phase, implying a Chern number $c=1$ in the gapless limit. Rigorously, the degeneracy between the FB and the dispersive band hinders a well-defined Chern number, but it is possible to open a gap at the singular point by introducing additional terms in the Hamiltonian. We will use the XYH lattice as an example to show that by including the spin-orbit coupling (SOC), the FB is attached with a well-defined Chern number identical to the LL \cite{wu2008orbital, zhang2011quantum}.

\begin{figure}[tbp]
\begin{center}
\tabcolsep=-0.5cm
\begin{tabular}{c}
\includegraphics[width=8.5cm]{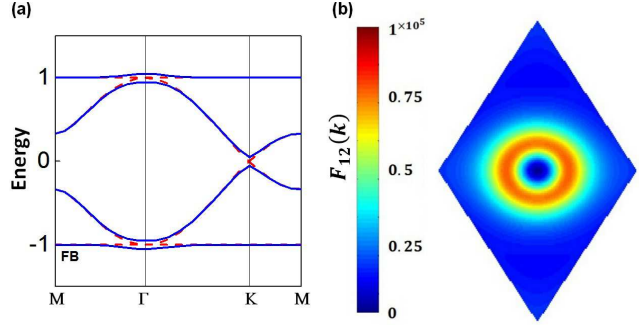}
\end{tabular}
\end{center}
\caption{(a) Band structure of the XYH lattice with SOC (blue solid curves)and without SOC (red dashed curves). $\lambda$ in Eq. (\ref{eq:Hsoc}) is set to be 0.05. (b) Berry curvature $F_{12}(\textbf{k})$ of the FB in the momentum space.} \label{fig:SOC_Berry}
\end{figure}

The atomic form of SOC is: $H_{SOC}=\lambda \hat{L}\cdot\hat{S}$, where $\hat{L}$ is the orbital anglular momentum operator and $\hat{S}$ is the spin-$\frac{1}{2}$ operator. For the $p$-orbitals, $H_{SOC}$ becomes:
\begin{eqnarray}
H_{soc}=\lambda(ic_{p_{x\downarrow}}^\dagger c_{p_{y\downarrow}}-ic_{p_{z\uparrow}}^\dagger c_{p_{y\uparrow}}+c_{p_{z\uparrow}}^\dagger c_{p_{x\downarrow}}-c_{p_{z\downarrow}}^\dagger c_{p_{x\uparrow}}\nonumber\\
+ic_{p_{z\uparrow}}^\dagger c_{p_{y\downarrow}}-ic_{p_z{\downarrow}}^\dagger c_{p_{y\uparrow}}+h.c.),
\end{eqnarray}
To further reduce $H_{SOC}$ into the $p_{xy}$ subspace, we keep the leading terms only:
\begin{eqnarray}
H_{soc}=\lambda(ic_{p_{x\downarrow}}^\dagger c_{p_{y\downarrow}}-ic_{p_{x\uparrow}}^\dagger c_{p_{y\uparrow}})+h.c.,
\end{eqnarray}
which do not include coupling between different spin components, so that the spin-up and spin-down spaces are independent. For each spin subspace, $H_{SOC}$ in the the XYH lattice can be written as a $4\times4$ matrix:
\begin{eqnarray} \label{eq:Hsoc}
H_{soc,\sigma=\pm}^0=\pm\lambda\left(
                                    \begin{array}{cccc}
                                      0 & -i & 0 & 0 \\
                                      i & 0 & 0 & 0 \\
                                      0 & 0 & 0 & -i \\
                                      0 & 0 & i & 0 \\
                                    \end{array}
                                  \right)
,
\end{eqnarray}
This SOC represents an imaginary on-site coupling at $A(B)$ sublattices, independent of momentum. After adding Eq. (\ref{eq:Hsoc}) to Eq. (\ref{eq:Hpxy}), we can solve the eigenvalue problem again and calculate the Berry curvature according to Eq. (\ref{eq:berryconn}). The result is shown in Fig. \ref{fig:SOC_Berry}. As expected, an energy gap is created by SOC at the degenerate point, which also induces dispersion of the FB around the splitting point [Fig. \ref{fig:SOC_Berry}(a)]. The Berry curvature exhibits an interesting ring pattern [Fig. \ref{fig:SOC_Berry}(b)]. The integration of the Berry curvature within BZ gives $c=1$.

To sum up, the FB and LL bear many similarities. They are dispersionless, macroscopically degenerate and topologically nontrivial. It is natural to speculate that the FB may spawn the same physics as the LL. The most interesting possibility is to realize the FQH state in the FB without the need of a strong magnetic field. Qi has recently shown a systematic approach to construct the FQH-state in the FB under a pseudo-Coulomb interaction, which rationalizes the speculation \cite{PRL11QiMap}. However, the readers should still be aware of some critical differences. Firstly, the LLs are fully gapped from each other, whereas the FB in the original from is degenerate with a dispersive band at some momentum point. A full gap is important to define a protected topology, as well as to prevent bands from mixing with each other when the Coulomb interaction is included. Secondly, although the degenerate point of FB can be removed by certain mechanisms, such as SOC, the band flatness is inevitably affected by creating the gap [Fig. \ref{fig:SOC_Berry}(a)], which partially destroys the kinetic quench of electrons, unfavorable for the FQH state. Finally, the Berry curvature $F_{12}\textbf(k)$ of FB is centered around the singular point, whereas that of LL is uniform and featureless. The same Chern number, i.e. the integration of $F_{12}\textbf(k)$, only ensures that the FB and the LL can be adiabatically connected at the single-electron level. However, it is not guaranteed that the FQH state on the LL can be adiabatic continued onto the FBs when the real Coulomb interaction is included. Therefore, to mimic the LL, the FB is expected to be (1) nearly flat; (2) well separated from the other bands; and (3) uniform in terms of Berry curvature \cite{CRP13ReviewFB}. However, these ideal conditions are usually difficult to satisfy simultaneously. It then relies on quantitative calculations to determine the existence of FQH state in the FB. As shown later in Sec. \ref{sec:phasediagram}, the FQH state appears only within a specific region of parameter space. This poses rather stringent criteria for material realization.

\section{Many-body phases} \label{sec:Manybody}

Let us now consider what will happen to the FB, when interactions are added. Since the FB contains macroscopic degeneracy, the Coulomb interactions, even small, becomes critical. In this Section, we show that various many-body phases emerge from the FB depending on the interaction type.

\subsection{Ferromagnetism}\label{sec:FM}

We first follow Tasaki's approach \cite{PTP98Tasaki} to show that ferromagnetism naturally arises on the FB as a consequence of onsite Coulomb interaction:
\begin{eqnarray}\label{eq:onsite}
H_{int}=U \sum_{i,a} \hat{n}_{i,a,\uparrow} \hat{n}_{i,a,\downarrow}
\end{eqnarray}

According to the Mermin-Wagner theorem \cite{PRL66MerminWagner}, for a 2D system there is no true long-range order at finite temperature. Hence our discussion should be constrained on \emph{finite-size} 2D lattices. The two extended eigenstates of the FB are thus absent and all the localized states ${\psi_{FB,\textbf{R}}}$ form a linear-independent and complete basis to construct many-body ground states within the FB subspace. In general, we can write the many-body state as:

\begin{eqnarray}\label{eq:PSI_general}
\Psi_{GS}=\sum_{\{\textbf{R}\}\{\textbf{R'}\}}f_{\{\textbf{R}\}\{\textbf{R'}\}}\prod_{\textbf{R}} \psi^\dag_{FB,\textbf{R}\downarrow} \prod_{\textbf{R'}} \psi^\dag_{FB,\textbf{R'}\uparrow} |0\rangle ,
\end{eqnarray}
where $\{\textbf{R}\}$ and $\{\textbf{R'}\}$ denote the sets assigned with the down and up spins, respectively. The sum runs over all configurations subject to an electron number $N_e$. $f_{\{\textbf{R}\}\{\textbf{R'}\}}$ is the linear combination coefficient.

Since each electron occupies one localized plaquette, energy cost occurs when two plaquettes carrying the opposite spins touch. This means that as long as possible, we demand $f_{\{\textbf{R}\}\{\textbf{R'}\}}=0$ if $\langle\psi_\textbf{R}|\psi_\textbf{R'}\rangle \neq 0$. For all the aforementioned FB models, we have one plaquette within each unit cell. Hence, this ``no double occupancy'' condition can be satisfied when $N_e \leq N_c$. To ensure that any two touching plaquettes carry the same spin, after selecting $N_e$ plaquettes out of the total $N_c$ unitcells, we group touching plaquettes as clusters ${C_\alpha}$ and assign the same spin for plaquettes within one cluster. Accordingly, the groundstate wavefunction can be written as:

\begin{eqnarray}\label{eq:PSIC}
\Psi_{GS}=\sum_{\{C_\alpha,\sigma_\alpha\}} f(\{C_\alpha, \sigma_\alpha\}) \prod_\alpha \prod_{\textbf{R}\in C_\alpha} \psi_{FB,\textbf{R},\sigma_\alpha}^\dag |0\rangle
\end{eqnarray}

When $N_e<N_c$, $\Psi_{GS}$ is highly degenerate. The special case is $N_e=N_c$, when all the plaquettes connect into one cluster. Equation (\ref{eq:PSIC}) is then reduced to:

\begin{eqnarray}\label{eq:PsiFM}
\Psi_{GS}=\sum_{\sigma=\uparrow,\downarrow} f_\sigma \prod_\textbf{R} \psi_{\textbf{R},\sigma}^\dag |0\rangle ,
\end{eqnarray}
which represents a fully-polarized ferromagnetic state apart from the trivial $(2S_{max}+1)$-fold degeneracy.

The FB ferromagnetism essentially arises from a percolation picture. When the electron filling is beyond a critical threshold, a unique infinite cluster forms, driving the system from paramagnetic to ferromagnetic. Recently, Maksymenko et al. have carefully studied this so-called ``Pauli-correlated'' percolation problem, revealing the details of the paramagnetic-ferromagnetic transition \cite{maksymenko2012flat}. In some sense, the FB ferromagnetism is complementary to Nagaoka's ferromagnetism \cite{PR66NagaokaFM}. These two types of ferromagnetism push the Stoner's criterion, $D(E_F)U>1$ to two extreme limits: Nagaoka's ferromagnetism takes place with an infinitely large $U$, while the FB provides an infinitely large density of states $D(E_F)$ at the Fermi level.

\subsection{Superconductivity}\label{sec:SC}

Since the FB gives rise to a divergence of density of states, it is intuitive that the pairing instability will become anomalous compared with the normal metal with a quadratic band dispersion. Miyahara et al. \cite{PhysC07BCSFB} considered an attractive interaction in the FB. A two-band BCS mean-field study was then performed, which showed that the BCS gap $\Delta(0)$ and the critical temperature $T_c$ were linear to the pairing potential $V_p$.

A simple demonstration can be made within the single band picture. Let us consider a BCS-type attractive potential between the FB states:
\begin{eqnarray}
H_{int}=-V_p\sum_{\textbf{k},\textbf{k'}} \psi^\dag_{FB,\textbf{k},\uparrow}\psi^\dag_{FB,\textbf{-k},\downarrow}\psi_{FB,\textbf{-k'},\downarrow}\psi_{FB,\textbf{k'},\uparrow}, \end{eqnarray}
the standard mean-field gap equation is (See, for example, Sec. 2.2 in \cite{Alex10SC}):
\begin{eqnarray}
\frac{1}{V_p}=\int d\xi \frac{D(\xi)}{\sqrt{\xi^2+\Delta(T)^2}} \tanh \frac{\sqrt{\xi^2+\Delta(T)^2}}{2k_B T},
\end{eqnarray}
where $\xi_{\textbf{k}}=\epsilon_{\textbf{k}}-\mu$. For the FB, $D(\xi)=\delta(0)$. Therefore,at zero temperature, we have:
\begin{eqnarray}
\Delta(0)=V_p
\end{eqnarray}
At $T_c$, we have:
\begin{eqnarray}
\frac{1}{V_p}=\frac{1}{2k_B T_c}
\end{eqnarray}
In summary,
\begin{eqnarray}
\Delta(0)=2k_B T_c =V_p
\end{eqnarray}
For comparison, a normal dispersive band leads to
\begin{eqnarray}
\Delta(0)\sim T_c \sim e^{-\frac{1}{D(E_F)V_p}}
\end{eqnarray}
Consequently, a much higher $T_c$ is expected in the FB.

\subsection{Wigner crystal}\label{sec:WC}
\begin{figure}[tbp]
\begin{center}
\tabcolsep=-0.5cm
\begin{tabular}{c}
\includegraphics[width=5cm]{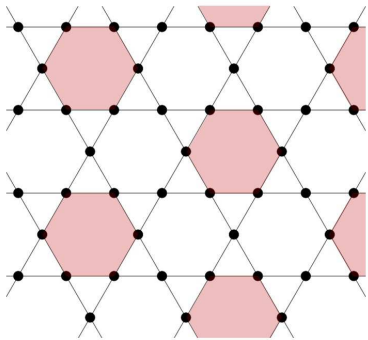}
\end{tabular}
\end{center}
\caption{The Wigner crystal phase in kagome lattice corresponding to the filling $N_e/N_c=1/3$ }\label{fig:wigner}
\end{figure}

Recall Eq. (\ref{eq:PSI_general}). For low electron filling without percolation, the ground-state configuration will be fixed by Coulomb interactions between distant sites. One can imagine that the occupied plaquettes should form special patterns in order to lower the interaction energy as much as possible. This is nothing but the Wigner crystal phase, which comes out when the potential energy dominates the kinetic energy. The FB is a natural platform to realize Wigner crystal, because the kinetic energy is fully quenched. Wu and Das Sarma have systematically studied the XYH lattice for various Wigner crystalline orders at different filling numbers \cite{PRL07Wu}. For a simple example, consider a NN Coulomb interaction in the kagome lattice:

\begin{eqnarray}
H_{int}=U' \sum_{\langle ia,jb \rangle} \hat{n}_{ia} \hat{n}_{jb},
\end{eqnarray}
where the summation runs over NN sites. The ground-state configuration of plaquettes without costing the NN Coulomb potential is depicted in Fig.\ref{fig:wigner} corresponding to the filling $N_e/N_c=1/3$.

\subsection{FQH state}\label{sec:FQH}

The FQH state on the LL is an important theme of modern condensed matter physics, which exhibits unusual features, such as fractionalization and entanglement, without any classical analogue \cite{RMP99FQH}. In Sec. \ref{sec:compare}, we have compared the FB with the LL in detail. If a similar state in the FB exists, the features of the FQH state should be reproduced. Regnault and Bernevig \cite{PRX11FCI} have suggested a comprehensive set of evidences accessible by numerical calculations to justify the FQH state in the FB. We summarize these evidences as follows.

\begin{enumerate}
\item Fractional Hall conductance. For a $1/n$-filled FB, the FQH state, if exists, features a Hall conductance equal to $1/n$ conductance quantum.  The many-body Hall conductance can be obtained by observing the evolution of the lowest-lying $n$-fold degenerate many-body energy levels under twisted boundary conditions \cite{PRB85NiuWu}. The levels evolve within the $n$-fold multiplet as the twisted phase increases, with a periodicity in $2n\pi$ to give the $1/n$ quantized Hall conductance. The Hall conductance can also be calculated from the many-body Chern number.
\item Fractionalized excitations. The quasiparticle excitations of a FQH state are very unique in the sense that they carry a fractional charge and exhibit the fractional statistics. The fractionally charged quasiparticles usually obey the generalized Pauli exclusion principle \cite{PRL01HaldaneStatis,PRL94WuFracStatis}, which places strong constraints on the precise number of low-lying excitation states \cite{PRL96WuFracStatis}. Numerically, the excitations can be introduced by changing the number of electrons in the system. The existence of FQH state in the FB can be justified by counting whether the total number of low-energy states in this excited system agrees with the exclusion principle.
\item Entanglement. A fundamental difference between the FQH state and the other many-body states mentioned in this Section is the electron correlation. The many-body wavefunction $\Psi_{GS}$ of the states discussed in Sec.\ref{sec:FM}, \ref{sec:SC} and  \ref{sec:WC} can all be written as a direct product of single-electron states [See Eq. (\ref{eq:PsiFM}) for example], whereas  $\Psi_{GS}$ of a FQH state is highly entangled \cite{PRL06WenTopoWave}. Numerically, the entanglement is shown by firstly constructing the density matrix $\hat\rho=|\Psi_{GS}\rangle\langle\Psi_{GS}|$. Then, the electrons are divided into two partitions A and B, and the degrees of freedom of partition B are traced out leaving a reduced density matrix $\hat\rho_A=Tr_B\hat\rho$. The reduced density matrix captures various properties of the ground-state entanglement, such as the entanglement spectrum \cite{PRL08HaldaneEntangSpectrum} and entanglement entropy \cite{PRL06EntangleEntropy}.
\end{enumerate}

\section{Phase diagram: a case study} \label{sec:phasediagram}

The variety of possible phases implies a rich phase diagram in the FB. Intuitively, ferromagnetism can be achieved most conveniently because it relies on the strongest on-site Coulomb repulsion. The BCS-type superconductivity requires an effective attractive interaction, which can be sought in materials with considerable electron-phonon coupling. There is competition between the Wigner crystal and FQH state in the FB, just like in the LL. When the FB is not exactly dispersionless, e.g. the case of Fig. \ref{fig:SOC_Berry}, the electrons may take advantage of the band dispersion by condensing into a trivial Fermi liquid. The many-body states survive only when the Coulomb interaction dominates the band dispersion. However, if the Coulomb interaction is larger than the band gap, additional dispersive bands will be mixed, which also destroys the FB physics.

To see this competition quantitatively, let us calculate an example \cite{PRB14WeiLi}. We will consider the XYH lattice. The single-electron part of Hamiltonian is given by Eq.(\ref{eq:Hpxy})+Eq.(\ref{eq:Hsoc}). For simplicity, we assume that the ferromagnetism presents as the result of on-site Coulomb interaction. Thus the spin-up and spin-down components are separated, and we focus on one of the spin component. This assumption effectively reduces the problem into a spinless one. The remaining Coulomb interaction up to NN then consists of:
\begin{eqnarray}
 H_{int} &=& U_0\sum_{i}\hat{n}_{i,p_x}\hat{n}_{i,p_y} + U_1\sum_{\langle i,j \rangle,\alpha}\hat{n}_{i,p_x}\hat{n}_{j,p_x}
\notag\\
&+& U'_1\sum_{\langle i,j \rangle}\hat{n}_{i,p_x}\hat{n}_{j,p_y}
\end{eqnarray}
Note that for convenience of writing the interaction terms, the index here is slightly different from Eq.(\ref{eq:hopping}). Namely, $i,j$ label the sites instead of the unit cells.

We consider a finite-size lattice with the bottom FB $\frac{1}{3}$ filled, e.g. $N_c=4\times6=24$,  $N_e=8$. Note that the XYH model has the particle-hole symmetry, so this filling can also be considered as the top FB $\frac{1}{3}$ empty. Since the Hilbert space is finite, we can numerically diagonalize the many-body Hamiltonian to get the eigenvalues and eigenstates, which is known as the exact diagonalization (ED) technique.

\begin{figure}[tbp]
\begin{center}
\tabcolsep=-0.5cm
\begin{tabular}{c}
\includegraphics[width=7cm]{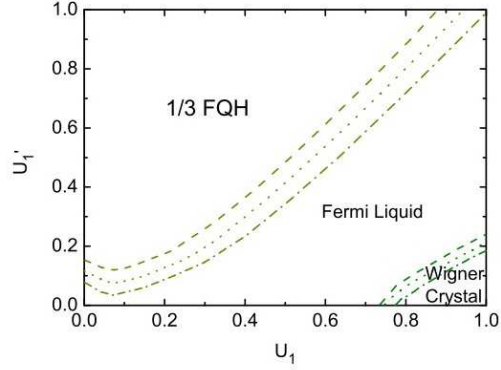}
\end{tabular}
\end{center}
\caption{Phase diagram \cite{PRB14WeiLi} including FQH state, Wigner crystal state and Fermi liquid in $U_1-U_1'$ plane at the $\frac{1}{3}$-filling with the system size $N_c=4\times6$. The dashed, dotted, and dashed-dotted lines represent the phase boundary for $U_0=0.3,0.4$ and $0.5$, respectively. }\label{fig:phase}
\end{figure}

Obviously, the electronic property is a function of $U_0$, $U_1$ and $U_1'$. Based on the ED results, a phase diagram is constructed (Fig.\ref{fig:phase}). Three phases, including FQH state, Wigner crystal and Fermi liquid, appear at different regions of the parameter space. The FQH state reigns when $U'_1$ dominates. For small $U'_1$, the ground state is at first a Fermi liquid, and then evolves into a Wigner crystal at sufficiently large $U_1$. In addition, if $U_0$ increases, the phase boundaries shift toward the larger value of $U_1$. It is interesting to notice that different interaction terms play very different roles in defining the many-body ground state, which provides a useful guidance to target a specific phase in real materials.

\begin{figure}[tbp]
\begin{center}
\tabcolsep=-0.5cm
\begin{tabular}{c}
\includegraphics[width=7cm]{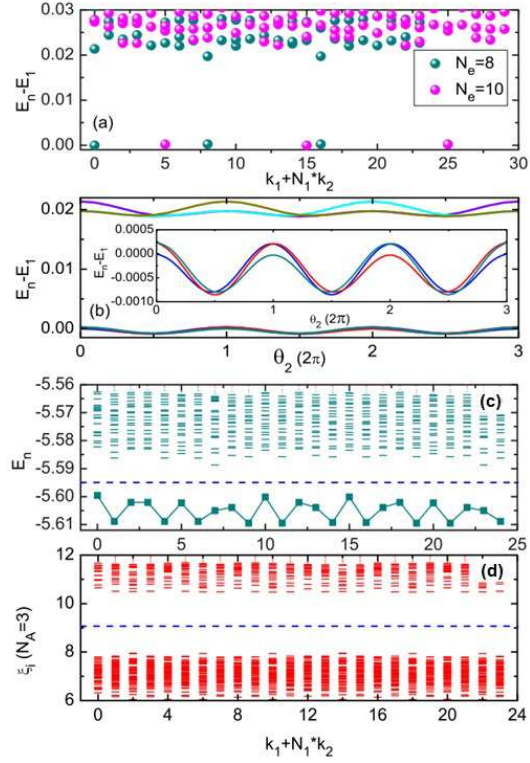}
\end{tabular}
\end{center}
\caption{(a) Low-energy spectrum for $N_c=4\times6, N_e = 8$ and
$N_c=5\times6, N_e=10$.(b) Evolution of the threefold degenerate ground state under twisted boundary conditions. (c)  Low-energy spectrum for $N_c=5\times5, N_e=8$, i.e. one electron less than the $\frac{1}{3}$ filling. (d) Particle entanglement spectrum for $N_c=4\times6, N_e = 8$. \cite{PRB14WeiLi}}\label{fig:ED}
\end{figure}

It is helpful to see how the ED results dictate the FQH features as described in Sec. \ref{sec:FQH}. Figure \ref{fig:ED}(a) shows the low-energy spectrum obtained from ED, which can be considered as a many-body counterpart of the single-electron band structure. The periodic boundary condition is employed:
\begin{eqnarray}
\Psi\{\textbf{r}_i+N_{c_{1(2)}}{\textbf{R}_{1(2)}}\}=\Psi\{{\textbf{r}_i}\},
\end{eqnarray}
where $\textbf{R}_{1(2)}$ is the primitive vector of the unit cell. With this translational symmetry, two good integer quantum numbers $(k_1,k_2)$ can be defined to reduce the Hamiltonian into specific momentum sectors $\textbf{q}=(2\pi k_1/N_{c_1}, 2\pi k_2/N_{c_2})$. The horizontal axis of Fig. \ref{fig:ED}(a) denotes different momentum sectors. The energy spectrum shows two necessary features of a $\frac{1}{3}$-FQH state. One is a three-fold degeneracy of the lowest energy levels; the other is a clear gap between this three-fold multiplet and the other levels. Both of these two features are absent for a Fermi liquid. Furthermore, Fig. \ref{fig:ED}(b) shows the evolution of the three-fold multiplet under twisted boundary conditions:
\begin{eqnarray}
\Psi\{\textbf{r}_i+N_{c_{1(2)}}{\textbf{R}_{1(2)}}\}= e^{i\theta_{1(2)}} \Psi\{{\textbf{r}_i}\}
\end{eqnarray}
The three states are found to evolve into each other with level crossing and to keep being separated from the other levels when $\theta_2$ increases . Eventually, it evolves back to the initial configuration when $\theta_2=6\pi$. This behavior indicates that the quantized
Hall conductance is $\sigma_H=\frac{1}{3}e^2/h$. Note that although a Wigner cyrstal may have a similar energy spectrum as the FQH-state, i.e. degeneracy and gap, the spectral flow is different. The FQH-state and the Wigner crystal can be further differentiated by the charge density distribution.

By adding one hole into the system, the quasihole excitation spectrum [Fig. \ref{fig:ED}(c)] reviews the fractional exclusion statistics. An energy gap is clearly visible in Fig. \ref{fig:ED}(c). According to the generalized Pauli principle of $\frac{1}{3}$-Laughlin state, the total number of states below the gap  is given by \cite{PRL01HaldaneStatis,PRX11FCI}:
\begin{eqnarray}\label{eq:holecount}
N = N_c \frac{(N_c-2N_e-1)!}{N_e ! (N_c-3N_e)!}
\end{eqnarray}
For $N_c=25$, $N_e=8$, Eq. (\ref{eq:holecount}) gives $N=25$. It is straightforward to check that there are indeed 25 states below the gap in Fig. \ref{fig:ED}(c).

In the end, following Regnault and Bernevig's recipe \cite{PRX11FCI}, we partition the electrons, and calculate the reduced density matrix $\hat\rho_A$ based on the ground-state wavefunction. If we define an  ``entanglement Hamiltonian'' $H_A$, we may write $\hat\rho_A=e^{-H_A}$, and the set of eigenvalues of $H_A$ constitutes the entanglement spectrum of the ground state. A clear entanglement gap is oberseved in the spectrum [Fig. \ref{fig:ED}(d)], which is a typical signature of FQH state. Also the counting of states below the gap satisfies the fractional statistics. All these numerical evidences confirm the existence of FQH state in the FB.

\section{Material realization} \label{sec:material}

The primary challenge to reproduce the FB in real materials is the degree of flatness, because the destructive interference can be easily destroyed in real materials by additional hopping processes as well as hybridization with other orbitals. In addition, even if a FB exists, it remains uncertain whether the Fermi level happens to be around it. For complex structures, there is no straightforward way to determine either of these two factors. Therefore, in general first-principles calculations \cite{RMP89DFT} are inevitable to identify possible candidates. In this section, we mention two promising directions for first-principles exploration. One is to search among the existing quantum frustrated materials. These materials were previously synthesized to access the quantum spin liquid state, but bear many similarities to the FB models. The other direction is to design organometallic frameworks thanks to the advances of nanotechnology and organic chemistry.

\subsection{Existing quantum frustrated materials}

An increasing number of quantum frustrated materials has been synthesized and measured, aiming at discovering the long-pursued quantum spin liquid state \cite{Science08PALeeQSL, Nature10BalentsQSL}. From the aspect of material sciences, it is interesting to notice that many of these materials have the exact lattice and/or orbital required by the FB models. For example, in Sec. \ref{sec:line}, we have discussed that the kagome lattice is a typical line graph to obtain the FB. In the field of quantum spin liquid, it is also widely used as the prototypical lattice to create spin frustration. In experiment, families of kagome antiferromagnets are known, such as jarosites, SCGO(x), volborthite and herbertsmithite (See \cite{Book11FrustMag} and references therein). The question is then whether we can seek for FBs in these existing materials.

\begin{figure}[tbp]
\begin{center}
\tabcolsep=-0.5cm
\begin{tabular}{c}
\includegraphics[width=7cm]{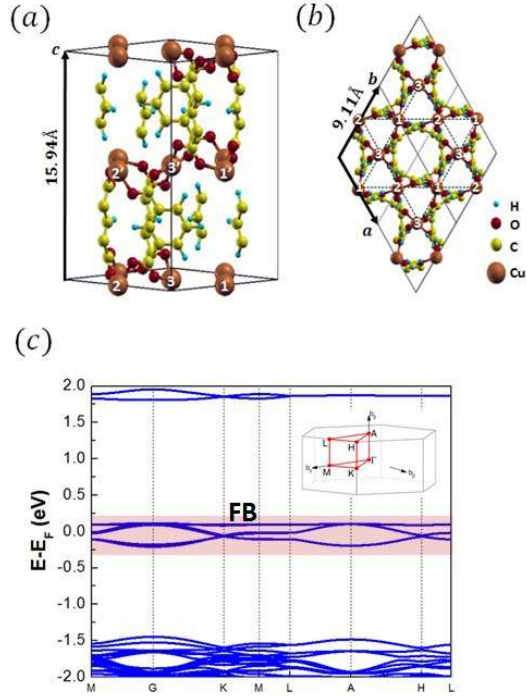}
\end{tabular}
\end{center}
\caption{(a) Side view and (b) top view of atomic structure of Cu(1,3-bdc). The white numbers on Cu atoms label the three inequivalent sites of a kagome plane. (c) First-principles band structure (within the spinless local density approximation) of Cu(1,3-bdc)}\label{fig:CuBDC}
\end{figure}

As an example, Fig. \ref{fig:CuBDC}(a) shows the crystal structure and first-principles bands of Cu(1,3-bdc) recently synthesized to achieve the quantum spin liquid state \cite{JACS08CuBDC}. The key structural feature is that the Cu atoms form layered kagome lattices (Fig. \ref{fig:CuBDC}(b)). Its single-electron band around the Fermi level evidently reproduces Fig. \ref{fig:kagome}(b) upside down, i.e. a negative $t$. The FB above the Fermi level is nearly completely flat. The reason is that around the Fermi level, there is effectively one $d_{x^2-y^2}$ orbital per Cu site, which is well separated from the other atomic orbitals due to the crystal field splitting. For other quantum frustrated materials containing the kagome lattice, the flatness could be much worse because of complicated orbital degrees of freedom. The main concern, however, is that Cu(1,3-bdc), like many other quantum frustrated materials, is in the Mott insulating regime, which corresponds to the half filling and a sufficiently large $U$. Strictly speaking, the band picture becomes invalid. It require significant doping to tune the material away from the Mott insulating phase, and to shift the Fermi level close to the FB.

An alternative way is to achieve a magnon FB in these materials which acts as a bosonic version of the electronic FB. Considering that the mapping from the Hubbard model to the Heisenberg spin model retains the structure of the hopping matrix $\{t_{ia,jb}\}$, the magnon FB is naturally expected. Specifically, it has been proposed that by fully polarizing the ground state with an external magnetic field, the low-energy magnetic excitation can be described by a boson field with the same hopping processes as the electron, which in the end produces a magnon FB \cite{schulenburg2002macroscopic,arxiv11MeiChiral}.

\subsection{Organometallic frameworks}

\begin{figure}[ht]
\includegraphics[width=6cm]{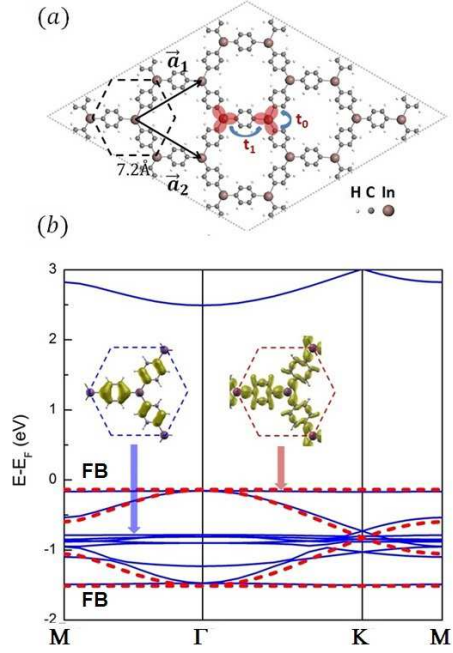}
\caption{ \label{fig:IPOF}(a) The atomic structure of IPOF. (b) Band structure without SOC from first-principles calculation (blue solid curves) and model Hamiltonian Eq. (\ref{eq:Hpxy}) (red dashed curves). Insets are the wavefunction isosurfaces of two states denoted by arrows. \cite{PRL13IPOF}}
\end{figure}

\begin{figure}[ht]
\includegraphics[width=0.5\textwidth]{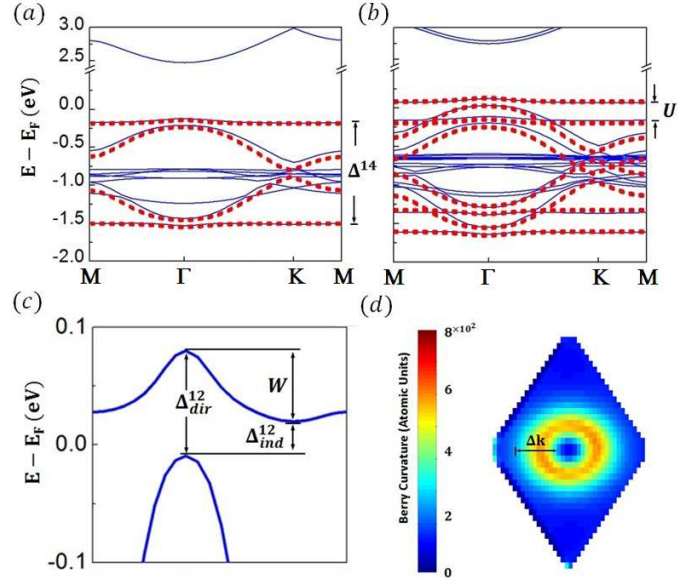}
\caption{\label{fig:DFTband}(a) Band structure with SOC. (b) Band structure with SOC when doping one hole into the unit cell. The solid (blue) curves are DFT results. The dashed (red) curves are from model Hamiltonian Eq. (\ref{eq:Hpxy})+Eq. (\ref{eq:Hsoc}) (c) Zoomed-in band plot around the Fermi level. (d) Berry curvature $F_{12}(\textbf{k})$ of the FB in IPOF from first-principles calculation. \cite{PRL13IPOF}}
\end{figure}

Advances on synthetic chemistry and nanotechnology have shown the potential in producing complex lattices \cite{AngW09Sakamoto,Nat05Barth}. Recent experiments using substrate-mediated self-assembly have successfully fabricated 2D organometallic frameworks with different lattice symmetry. These covalent organic frameworks are also found to exhibit remarkable thermal stability. Based on first-principles calculations, we have predicted topological band structures in many organometallic frameworks containing heavy metal atoms \cite{NatComm13Zhengfei,NanoLett13Zhengfei,PRL13Zhengfei,PRL13IPOF}.

To design an organometallic framework containing FBs, the basic strategy is to first identify proper molecular building blocks, and then link these blocks into the correct lattice geometry required by the FB model. The stability of the designed structure should be carefully examined, e.g. by first-principles lattice relaxation and phonon calculations.

In 2013, we proposed a first-principles design to reproduce Eq. (\ref{eq:Hpxy}) in a 2D Indium-Phenylene Organometallic Framework (IPOF) \cite{PRL13IPOF}.  The building block is the triphenyl-indium $In(C_6H_5)_3$, a common indium compound \cite{Book85Indium}.  The In atoms are linked by the phenylenes into a hexagonal lattice [Fig. \ref{fig:IPOF}(a)]. The resulting electronic band structure and wavefunction are summarized in Fig. \ref{fig:IPOF}(b). The top FB lies right below the Fermi level.  After including SOC in the calculation, the FBs become separated from the dispersive bands as predicted by the model study [Fig. \ref{fig:DFTband}(a)].

The FB ferromagnetism is shown within the first-principles formalism by manually reducing the number of valence electrons in the unit cell, while maintaining the charge neutrality with a compensating homogeneous background charge. This makes the top FB partially filled. Calculations reveal a spontaneous spin polarization. Figures \ref{fig:DFTband}(b) and \ref{fig:DFTband}(c) shows the band structure of this ferromagnetic ground state under doping. The spin-up and spin-down bands are separated apart, with the Fermi level shifting below the topmost spin-polarized FB.

The topology of the topmost spin-polarized FB is examined by directly calculating its Chern number based on its DFT wavefunctions and the edge property. The distribution of Berry curvature is shown in Fig. \ref{fig:DFTband}(d), in good agreement with Fig. \ref{fig:SOC_Berry}(b). The integration of the Berry curvature in the whole Brillouin zone gives $c=1$. Both the band structure and the Berry curvature can be nicely fitted to the XYH model. The first-principles results quantify several key energy scales associated with the FB. The results for IPOF are summarized in Tab. I. It is worth mentioning that first-principles calculations have also revealed the existence of FBs as described by Eq. (\ref{eq:Hkagome}) in an experimentally made organometallic framework \cite{NanoLett13Zhengfei,JACS13Kagome}.

\begin{table}[ht]
\caption{Energy scales associated with the FB in IPOF} \label{table}
\begin{ruledtabular}
\begin{tabular}{cccc}
  Property & Symbol & Value & Ref. \\
  \hline
  Band width & $W$ & $60meV$ & Fig.\ref{fig:DFTband}(c) \\
  Spin splitting & $U$ & $100meV$ & Fig.\ref{fig:DFTband}(b) \\
  Energy gap & $\Delta^{12}_{dir}$ & $90meV$  & Fig.\ref{fig:DFTband}(c) \\
   & $\Delta^{12}_{ind}$ & $30meV$ & Fig.\ref{fig:DFTband}(c) \\
  & $\Delta^{14}$ & $1.4eV$ & Fig.\ref{fig:DFTband}(a) \\
\end{tabular}
\end{ruledtabular}
\end{table}

\section{Summary and outlook}\label{sec:end}
Whenever an unconventional band structure is brought out from a conceptual model into a real-world material, a wide range of technological innovations will be triggered. The well-known examples are graphene \cite{RMP09GeimGraphene,RMP11Geim} and topological insulators \cite{RMP10Hasan,RMP11Qi}, whose band structures are featured with linear dispersion and nontrivial topology, respectively. In this article, we have reviewed the fascinating idea of FB, which combines together unusual dispersion and nontrivial topology, leading to even more exotic physics. The current studies have sketched a blueprint, but much more efforts are required to build the mansion. Theoretically, it is worth exploring other possibilities in the FB, such as higher Chern number \cite{PRB12HighChern,PRB12ArbiChern,PRB12ChernTwo}, other classes of topology \cite{PRB11LuSymProt,PRL12NonAbel} and higher dimensions \cite{PRB12FB3D, PRL13WuHighDim}. On the other hand, schemes to utilize the FB for practical applications are highly demanded. In the context of either spintronics or quantum computing, detailed discussions on characterizing and manipulating the electrons in the FB are still lacking. After all, the central task is undoubtedly to discover FBs in real materials, which may rely on close collaborations with chemists and material scientists. By bringing together researchers working on diverse aspects, a fruitful journey is expected to reach this romantic flatland.

\section{Acknowledgement}
Z.L. would like to thank X.-G. Wen and J.-W. Mei for patiently explaining their pioneering work on the FB. We appreciate the conversations with D.-N. Sheng, X.-L. Qi, Z.-C. Gu, C.-J. Wu, H. Katsura and W. Li, which help deepen our understandings on various FB-related topics. We also would like to express our gratitude to Miao Zhou for proofreading the manuscript. F.L. and Z.L. acknowledge supports from DOE-BES (DE-FG02-03ER46027). Y.S.W. acknowledges supports from the U.S. NSF Grant No. PHY-1068558.

\bibliography{review}

\end{document}